\documentstyle[sprocl]{article}

\def\be{\begin{equation}}
\def\ee{\end{equation}}
\def\bea{\begin{eqnarray}}
\def\eea{\end{eqnarray}}
\begin{document}

\title{CURRENT ISSUES IN HEAVY QUARK PRODUCTION}
\author{J. SMITH}

\address{C.N. Yang Institute for Theoretical Physics,
State University of New York at Stony Brook,
New York 11794-3840, USA}


\maketitle\abstracts{
We discuss heavy quark production in deep inelastic scattering (DIS).}

%\subsection{Using Other Word-Processing Packages}\label{subsec:wpp}

Consider the effect of a charm quark (with mass $m$) 
being produced in neutral current DIS. 
For simplicity we concentrate on virtual photon
exchange and neglect the bottom quark. 

Precise data are now available on $F_2(x, Q^2)$. 
In pQCD the operator product expansion and 
asymptotic freedom allow one to calculate the scaling violations. 
This is incorporated in the factorization theorem where short
distance effects are separated into coefficient functions and long
distance effects into parton densities (for lowest twist terms), 
whose evolution is governed by the DGLAP equations. 
A mass factorization scale $\mu$ is thereby introduced.
The corrections to such an analysis are of order $\Lambda^2/{Q^2}$. 
The coefficient and splitting functions have been calculated 
to order $\alpha_s^2$ with massless quarks. The first
problem is how the presence of a massive c quark affects the above analysis?
In principle there should not be any problem because massless quarks were only
used for calculational convenience. At small scales 
$\mu \ll m$ the three flavour splitting functions and 
parton densities should remain the same. The coefficient functions
then contain terms in $m$. The c quark contribution in DIS 
$F_{2,c,{\rm EXACT}}(x,Q^2,m^2)$ can then be calculated in this 
fixed three-flavour number scheme
with parton densities for u,d and s quarks (and antiquarks) and added to
the light quark component $F_{2,{\rm LIGHT}}(x,Q^2)$ 
to form $F_2(x,Q^2)$. 

As $Q^2$ increases large terms 
in $\ln(Q^2/m^2)$ develop which should be resummed to give stable
predictions. This means that for $Q^2 \gg m^2$ the c quark becomes effectively
massless and should be removed from the coefficient functions into the
light parton densities. When this has been done
there are four-flavour densities (including
one for a c quark) and four-flavour massless coefficient functions.
$F_{2,{\rm ZM-VFNS}}(x,Q^2)$ is the zero-mass variable flavour number scheme 
representation for DIS. 
Since the evolution in the DGLAP equations
covers all scales there should also be a variable flavour number scheme (VFNS)
which interpolates between the two descriptions above. Clearly the 
schemes are complementary in that there are overlap regions 
where one scheme is as
good as another and maybe much simpler to use, but one has to be explicit in
the construction of a VFNS by specifying which parton densities and
which coefficient functions are used. 
The ACOT scheme is an order $\alpha_s$ VFNS and recently
a simplified version of it has been proposed \cite{kos}. 
The CTEQ5 parton density set \cite{cteq5} contains both
fixed flavour densities and VFNS sets which use the ACOT scheme.

What is known in higher order pQCD? In \cite{bmsn1} we established
an all-order relation between $F_{2,{\rm ZM-VFNS}}$ and 
$F_{2,c,{\rm EXACT}}$ which implied an all-order 
relation between light and heavy quark densities.
The operator matrix elements containing the terms in powers 
of $\ln(\mu^2/m^2)$, which should be incorporated into boundary
conditions on the evolution of the four-flavour densities (including a
c quark density), were calculated to order $\alpha_s^2$. 
Starting from a three-flavour set of densities
in \cite{grv98} we have constructed a set of four-flavour densities 
in \cite{chsm} which satisfy these boundary conditions and resum the 
large logarithms. Each set has a different gluon density and 
satisfies the momentum sum rule.  Thorne and Roberts \cite{thro}
tried to construct coefficient functions for use with such  
ZM-VFNS parton densities but concluded that the scheme was too complicated. 
They then proposed a different scheme based
on continuity of $dF_2(x,Q^2)/d\ln(Q^2)$ across flavour thresholds. 
This TR scheme has been implemented in order $\alpha_s$ in the MRST98 
\cite{mrst98} (and subsequent MRST) parton density sets. 

Two VFNS's now exist in order $\alpha_s^2$. One is 
called the CSN scheme \cite{csn1}. It uses heavy quark coefficient 
functions and begins with a term in order $\alpha_s^0$. 
The second is called the BMSN scheme \cite {csn1} \cite{bmsn2} and
is based on massless coefficient functions. 
The differences between all the VFNS's can be attributed 
to three ingredients entering their construction. The first one is the
mass factorization procedure carried out before the large logarithms can be
resummed. The second one is the matching condition imposed on the 
c-quark density, which has to vanish in the threshold region of the production
process. The third one is the construction of both three-flavour
and four-flavour sets of parton densities,
respecting the relations in \cite{bmsn1}.
Although designed to calculate $F_2(x,Q^2)$ 
both the BMSN and CSN schemes divide it into two pieces called 
$F_{2,c}(x,Q^2,\Delta)$ and $F_{2,{\rm LIGHT}}(x,Q^2,\Delta)$
respectively. The quantity $\Delta$ separates the collinear $c-\bar c$ pair 
production from the noncollinear pair production and is the value 
of the invariant mass above which they can be detected. This split 
makes $F_{2,c}(x,Q^2,\Delta)$ collinear safe. The BMSN
and CSN schemes are designed so they yield the same
result as the three-flavour NLO expression for $F_{2,c}(x,Q^2,\Delta)$ 
at $Q^2 = m^2$. In this region the four-flavour scheme fails. 
As $Q^2$ increases there are differences between the BMSN, CSN and 
three-flavour schemes reflecting both how the higher order terms 
are resummed into densities and the influence of terms in $m^2/Q^2$. 
At large $Q^2$ both the BMSN and CSN results for $F_{2,c}(x,Q^2,\Delta)$
agree numerically with the four-flavour results. 
For $x\approx 10^{-3}$ and $Q^2=10^3$ $({\rm GeV/c})^2$ 
both disagree by approximately ten percent with the three-flavour result 
showing that the resummation of the large logarithms is not a 
dramatic effect once one gets to order $\alpha_s^2$. 
The differences increase at larger $x$ which, for fixed $Q^2$,
is closer to the threshold at $Q^2(x^{-1}-1) = 4m^2$ 
because effects due to terms in $m$ are enhanced.

Note that we have designed the soft/collinear split
above without introducing a fragmentation function. 
We can do this for suitably defined inclusive quantities.
The next problem is the calculation of the single
particle inclusive rate for producing $D^*\,(\bar D^*)$ mesons. 
Every experimental measurement must specify thresholds 
(or regions in phase space) below which the mesons cannot be detected. 
Under these circumstances
potential collinear divergences remain in the final state. 
Is it better to keep the c-quark massive to avoid them or
to take the c-quark massless and absorb the singularities into 
fragmentation functions (and resum them via evolution)? 
Either way we need additional information on the parametrization 
of the fragmentation functions describing the transition 
from quarks to mesons. 
Also there is another scale $p_t$ and a new set of logarithms. 
Which logarithms are more important? 
There is no gain in resumming terms
in $\ln(Q^2/m^2)$ into either the parton densities 
or the fragmentation functions if they are not dominant. 
Does one need a VFNS for single particle inclusive production
or only a fixed flavour scheme and where
do they provide a better description of the data?

To begin we assume the differential production  
rate is measured at small $Q^2$ and is modelled by extrinsic production.
The NLO exclusive program HVQDIS \cite{hs}
uses three-flavour parton densities in the initial state and a massive c-quark
to predict the rate.
The reason for this is that potential collinear singularities 
involving the final state c-quarks are now controlled by $m$. 
We assumed a simple Peterson et al.,
type fragmentation function to model the transition from the 
$c (\bar c)$-quark to the
$D^*\,(\bar D^*)$ meson. The final parameters are the 
pole mass $m$, the fragmentation parameter 
$\epsilon$, the scale $\mu$, which
appears in the running coupling (and the parton densities) and
the probability for $c\rightarrow D^*$. The differential
rates for the $D^*\,(\bar D^*)$ mesons measured in the 
H1 \cite{H1} and ZEUS \cite{ZEUS}
experiments are in good agreement with this NLO calculation. 
A recent comparison is available in \cite{harris}.
The experimental groups have used HVQDIS to integrate over all phase space
to produce a quantity called $F_{2,c{\bar c}}(x,Q^2)$, 
the charm component of $F_2(x,Q^2)$.
However note this is based on the HVQDIS program, which does not
include diffractive production, intrinsic charm or resummation effects
at large and/or small $x$. It is also not the
same quantity we defined above with the parameter $\Delta$. 
Authors who claim agreement between their theory
and this experimental result should be aware of this fact.
HVQDIS has been used to produce
distributions in the invariant mass of the $c-\bar c$ 
(or $D^*-\bar D^*)$ pair and experimental results should be available soon.

The NLO calculation is complicated. Can one fit the same 
differential distributions and/or the experimental
$F_{2,c{\bar c}}(x,Q^2)$ 
in a four-flavour scheme with a massless c-quark? 
This can be tried at two levels. First can one calculate the single 
particle inclusive distributions with massless quarks? 
Most of the $D^*\,(\bar D^*)$
data is at small $p_t$ so the only large scale is $Q^2$ (and
$\mu^2$). Therefore one needs to understand how to partition the potentially 
large terms in $\ln(Q^2/\mu^2)$ and $\ln(\mu^2/m^2)$
between the parton densities and fragmentation
functions leaving a collinear safe partonic differential 
coefficient function.  Then how about $F_{2,c{\bar c}}(x,Q^2)$? 
Can one integrate over the final states to remove the
fragmentation function and give a four-flavour description at the level of the
coefficient functions? The latter investigation is more involved because every
time we integrate over a variable we encounter a potential new collinear 
singularity as $m\rightarrow 0$. These questions remain to be investigated.

We end with the comment that the all-orders proof of factorization 
in DIS including heavy quark effects given by Collins \cite{acot}  contains a
unitarity sum over {\it all} the final states and he therefore never defines
a collinear safe charm contribution. Also the advantage of using massless 
coefficient functions in VFNS's was first proposed in \cite{bmsn2}.

\section*{Acknowledgements}
This work was supported in part by NSF PHY-9722101. I would like to
thank B. Harris, E. Laenen and W.L. van Neerven for comments on
this report.

\end{document}